\documentclass[journal,12pt,onecolumn]{IEEEtran}

\usepackage{setspace}
\ifCLASSOPTIONonecolumn \doublespace \fi
\usepackage[latin1]{inputenc}
\usepackage{amsmath}
\usepackage{amsfonts}
\usepackage{amssymb}
\usepackage{graphicx}
\usepackage{cite}
\usepackage{fancyhdr}

\ifCLASSOPTIONonecolumn
\title{Reduced Complexity Decoding for Bit-Interleaved Coded Multiple Beamforming with Constellation Precoding}
\author{
\IEEEauthorblockN{Boyu Li\authorrefmark{1}, Hong Ju Park\authorrefmark{2}, and Ender Ayanoglu\authorrefmark{1}} \\
\IEEEauthorblockA{\authorrefmark{1}Center for Pervasive Communications and Computing, \\ Department of Electrical Engineering and Computer Science, \\ The Henry Samueli School of Engineering, \\ University of California, Irvine, \\ Irvine, California 92697-2625, \\ Email: boyul@uci.edu, ayanoglu@uci.edu} \\
\IEEEauthorblockA{\authorrefmark{2}Samsung Electronics, Suwon, Korea, honspark@samsung.com}
}
\date{} 
\else
\title{Reduced Complexity Decoding for Bit-Interleaved Coded Multiple Beamforming with Constellation Precoding}
\author{
\IEEEauthorblockN{Boyu Li\authorrefmark{1}, Hong Ju Park\authorrefmark{2}, and Ender Ayanoglu\authorrefmark{1}}
\IEEEauthorblockA{\authorrefmark{1}Center for Pervasive Communications and Computing, \\ Department of Electrical Engineering and Computer Science, \\ The Henry Samueli School of Engineering, \\ University of California, Irvine, \\ Irvine, California 92697-2625, \\ Email: boyul@uci.edu, ayanoglu@uci.edu}
\IEEEauthorblockA{\authorrefmark{2}Samsung Electronics, Suwon, Korea, honspark@samsung.com}
}
\date{} 
\fi

\begin{document}
\maketitle
\thispagestyle{fancy}

\ifCLASSOPTIONonecolumn
 \setlength\arraycolsep{4pt}
\else
 \setlength\arraycolsep{2pt}
\fi

\begin{abstract}

Multiple beamforming is realized by singular value decomposition of the channel matrix which is assumed to be known to both the transmitter and the receiver. Bit-Interleaved Coded Multiple Beamforming (BICMB) can achieve full diversity as long as the code rate $R_c$ and the number of employed subchannels $S$ satisfy the condition $R_cS \leq 1$. Bit-Interleaved Coded Multiple Beamforming with Constellation Precoding (BICMB-CP), on the other hand, can achieve full diversity without the condition $R_cS \leq 1$. However, the decoding complexity of BICMB-CP is much higher than BICMB. In this paper, a reduced complexity decoding technique, which is based on Sphere Decoding (SD), is proposed to reduce the complexity of Maximum Likelihood (ML) decoding for BICMB-CP. The decreased complexity decoding achieves several orders of magnitude reduction, in terms of the average number of real multiplications needed to acquire one precoded bit metric, not only with respect to conventional ML decoding, but also, with respect to conventional SD.  
\end{abstract}

\begin{IEEEkeywords}
MIMO, Beamforming, Constellation Precoding, Bit-Interleaved Coded Modulation, SD, Decoding Complexity.
\end{IEEEkeywords}

\ifCLASSOPTIONonecolumn \newpage \fi

\section{Introduction} \label{sec:Introduction}

Beamforming is employed in a Multi-Input Multi-Output (MIMO) system to achieve spatial multiplexing\footnotemark \footnotetext{In this paper, the term ``spatial multiplexing" is used to describe the number of spatial subchannels, as in \cite{Paulraj_ST}. Note that the term is different from ``spatial multiplexing gain" defined in \cite{Zheng_DM}.} and thereby increase the data rate, or to enhance the performance, when channel state information is available at the transmitter \cite{Jafarkhani_STC}. A set of beamforming vectors is obtained by Singular Value Decomposition (SVD) which is optimal in terms of minimizing the average Bit Error Rate (BER) \cite{Palomar_JTRBD}.

\IEEEpubidadjcol

It is known that an SVD subchannel with larger singular value provides greater diversity gain. Spatial multiplexing without channel coding results in the loss of the full diversity order \cite{Sengul_DA_SMB}. To overcome the diversity order degradation of multiple beamforming, Bit-Interleaved Coded Multiple Beamforming (BICMB) was proposed \cite{Akay_BICMB}, \cite{Akay_On_BICMB}. BICMB can achieve the full diversity order offered by the channel as long as the code rate $R_c$ and the number of subchannels used $S$ satisfy the condition $R_c S \leq 1$ \cite{Park_DA_BICMB}.

Bit-Interleaved Coded Multiple Beamforming with Constellation Precoding (BICMB-CP) converts a symbol into a precoded symbol and distributes it over subchannels \cite{Park_BICMB_CP}. The addition of the constellation precoder to BICMB, whose code rate $R_c$ is greater than $1/S$, provides the full diversity when the subchannels for transmitting the precoded symbols are properly chosen. However, BICMB-CP causes increased decoding complexity compared to BICMB. 

In this paper, Sphere Decoding (SD) with initial radius acquired by Zero-Forcing Decision Feedback Equalization (ZF-DFE) is used to calculate bit metrics of precoded symbols. The initial radius calculated by ZF-DFE \cite{Han_SD_IR}, which is also the metric weight of the Baiba point \cite{Agrell_CPS}, ensures no empty spheres. Based on SD, two techniques are applied to reduce the number of executions carried out by SD and the computational complexity of each SD execution, respectively. Conventional SD substantially reduces the complexity, in terms of the average number of real multiplications needed to acquire one precoded bit metric, compared with exhaustive search. With the techniques proposed in this paper, further reductions of orders of magnitude are achieved. The reduction becomes larger as the constellation precoder dimension and the constellation size increase.

The remainder of this paper is organized as follows: In Section \ref{sec:System_model}, the description of BICMB-CP is given. In Section \ref{sec:Decoding}, a reduced complexity decoding technique for BICMB-CP is proposed. In Section \ref{sec:Results}, complexity comparisons for different constellation precoder dimensions or modulation schemes are presented. Finally, a conclusion is provided in Section \ref{sec:Conclusion}.

\textbf{Notation:} Let $\textrm{diag}[\mathbf{B}_1, \cdots, \mathbf{B}_P]$ stand for a block diagonal matrix with matrices $\mathbf{B}_1, \cdots, \mathbf{B}_P$, and let $\textrm{diag}[b_1, \cdots, b_P]$ be a diagonal matrix with diagonal entries $b_1, \cdots, b_P$. The superscripts $(\cdot)^H$, $(\cdot)^T$, and $\bar{(\cdot)}$ stand for conjugate transpose, transpose and binary complement, respectively. Let $\mathbb{R}^+$ and $\mathbb{C}$ stand for the set of positive real numbers and complex numbers, respectively. Finally, let $N_t$ and $N_r$ stand for the number of transmit and receive antennas, respectively.

\section{BICMB-CP Overview} \label{sec:System_model}

Fig. \ref{fig:system_model} represents the structure of BICMB-CP. First, the convolutional encoder with code rate $R_c = k_c/n_c$, possibly combined with a perforation matrix for a high rate punctured code \cite{Haccoun_PCC}, generates the codeword $\mathbf{c}$ from the information bits. Then, the spatial interleaver distributes the coded bits into $S\leq\min(N_t, N_r)$ streams, each of which is interleaved by an independent bit-wise interleaver $\pi$. The interleaved bits are modulated by Gray mapped square QAM onto the complex-valued symbol sequence $\tilde{\mathbf{X}} = [\tilde{\mathbf{x}}_1 \, \cdots \, \tilde{\mathbf{x}}_K]$, where $\tilde{\mathbf{x}}_k$ is an $S \times 1$ complex-valued symbol vector at the $k^{th}$ time instant. It is assumed that each stream employs the same $2^M$-QAM constellation, where $M$ is the number of bits labeling a complex-valued scalar symbol. Let $\tilde{\chi} \subset \mathbb{C}$ of size $|\tilde{\chi}| = 2^M$ denote the complex-valued signal set of the square QAM.

\ifCLASSOPTIONonecolumn
\begin{figure}[!m]
\centering \includegraphics[width = 0.6\linewidth]{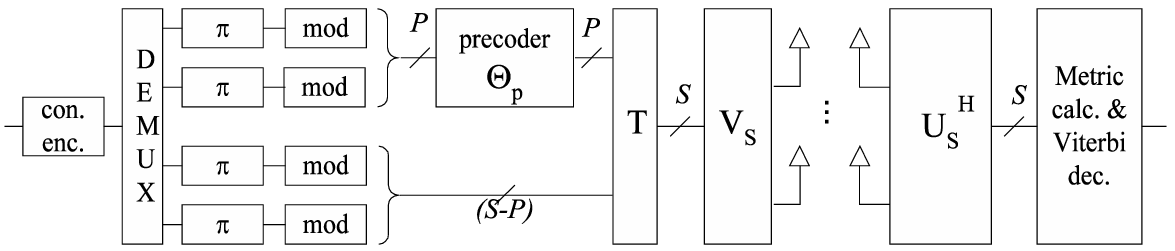}
\caption{Structure of BICMB-CP.} \label{fig:system_model}
\end{figure}
\else
\begin{figure}[!t]
\centering \includegraphics[width = 1.0\linewidth]{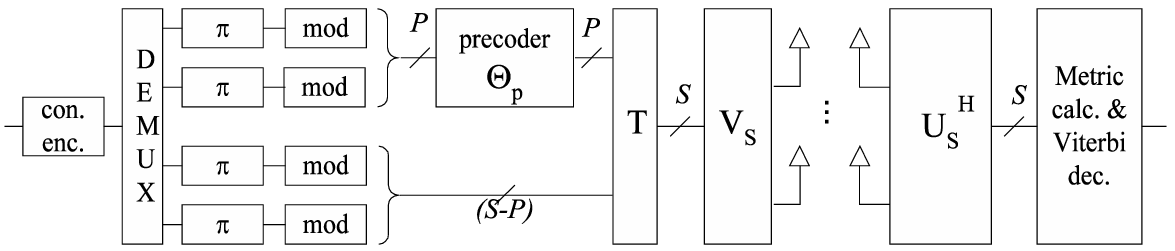}
\caption{Structure of BICMB-CP.} \label{fig:system_model}
\end{figure}
\fi

The complex-valued symbol vector $\tilde{\mathbf{x}}_k$ is multiplied by the $S \times S$ precoder $\boldsymbol{\Theta}$, which is defined as
\begin{align}
\mathbf{\Theta} = \left[ \begin{array}{cc} 
\mathbf{\Theta}_p & \mathbf{0} \\
\mathbf{0} & \mathbf{I}_{S-P}
\end{array} \right]
\label{eq:precoder_def}
\end{align}
where $\mathbf{\Theta}_p$ is the $P \times P$ unitary constellation precoding matrix that precodes the first $P$ modulated
entries of $\tilde{\mathbf{x}}_k$. The system is called Bit-Interleaved Coded Multiple Beamforming with Full Precoding (BICMB-FP) when all of the $S$ modulated entries are precoded, otherwise it is called Bit-Interleaved Coded Multiple Beamforming with Partial Precoding (BICMB-PP). The symbol generated by $\boldsymbol{\Theta}$ is multiplied by $\mathbf{T}$, which is an $S \times S$ permutation matrix, to map the precoded and non-precoded symbols onto the predetermined subchannels. Let us define $\mathbf{b}_p = \left[ b_p(1) \, \cdots \, b_p(P) \right]$ as a vector whose element $b_p(u)$ is the subchannel on which the precoded symbols are transmitted, and ordered increasingly such that $b_p(u) < b_p(v)$ for $u < v$. In the same way, $\mathbf{b}_n = \left[ b_n(1) \, \cdots \, b_n(S-P) \right]$ is defined as an increasingly ordered vector whose element $b_n(u)$ is the subchannel which carries the non-precoded symbols.

The MIMO channel $\mathbf{H} \in \mathbb{C}^{N_r \times N_t}$ is assumed to be quasi-static, Rayleigh, and flat fading, and perfectly known to both the transmitter and the receiver. Assume that the channel coefficients remain constant for a block of $K$ symbols. The beamforming vectors are determined by the SVD of the MIMO channel, i.e., $\mathbf{H} = \mathbf{U \Lambda V}^H$ where $\mathbf{U}$ and $\mathbf{V}$ are unitary matrices, and $\mathbf{\Lambda}$ is a diagonal matrix whose $s^{th}$ diagonal element, $\lambda_s \in \mathbb{R}^+$, is a singular value of $\mathbf{H}$ in decreasing order. When $S$ scalar symbols are transmitted at the same time, then the first $S$ vectors of $\mathbf{U}$ and $\mathbf{V}$ are chosen to be used as beamforming matrices at the receiver and the transmitter, respectively. In Fig. \ref{fig:system_model}, $\mathbf{U}_S$ and $\mathbf{V}_S$ denote the first $S$ column vectors of $\mathbf{U}$ and $\mathbf{V}$ respectively.

The spatial interleaver arranges the complex-valued symbol vector as $\tilde{\mathbf{x}}_k^\prime = [(\tilde{\mathbf{x}}_k^p)^T \, \vdots \, (\tilde{\mathbf{x}}_k^n)^T]^T = [\tilde{x}_{k,b_p(1)} \, \cdots \,
\tilde{x}_{k,b_p(P)} \, \vdots$ $\, \tilde{x}_{k,b_n(1)} \, \cdots \, \tilde{x}_{k, b_n(S-P)}]^T$, where $\tilde{\mathbf{x}}_k^p$ and $\tilde{\mathbf{x}}_k^n$ are the modulated entries to be transmitted on the subchannels specified in $\mathbf{b}_p$ and $\mathbf{b}_n$, respectively. Then, the $S \times 1$ received complex-valued symbol vector at the $k^{th}$ time instant $\tilde{\mathbf{r}}_k = [ (\tilde{\mathbf{r}}_{k}^p)^T \, \vdots \, (\tilde{\mathbf{r}}_{k}^n)^T]^T = [\tilde{r}_{k,1} \, \cdots \, \tilde{r}_{k, P} \, \vdots\, \tilde{r}_{k, P+1} \, \cdots \, \tilde{r}_{k,S}]^T$  is
\begin{align}
\tilde{\mathbf{r}}_k = \boldsymbol{\Gamma} \mathbf{\Theta} \tilde{\mathbf{x}}_k^\prime + \tilde{\mathbf{n}}_k, \label{eq:detected_symbol}
\end{align}
where $\boldsymbol{\Gamma} = \textrm{diag}[\boldsymbol{\Gamma}_p ,\boldsymbol{\Gamma}_n]$ is a block diagonal matrix, with diagonal matrices $\boldsymbol{\Gamma}_p = \textrm{diag}[\lambda_{b_p(1)}, $ $\, \cdots, \, \lambda_{b_p(P)}]$ and $\boldsymbol{\Gamma}_n = \textrm{diag}[\lambda_{b_n(1)}, \, \cdots, \, \lambda_{b_n(S-P)}]$, and $\tilde{\mathbf{n}}_k = [ (\tilde{\mathbf{n}}_{k}^p)^T \, \vdots \, (\tilde{\mathbf{n}}_{k}^n)^T]^T = [\tilde{n}_{k,1} \, \cdots \, \tilde{n}_{k, P} \, \vdots \, \tilde{n}_{k, P+1} \, \cdots \, \tilde{n}_{k,S}]^T$ is a complex-valued additive white Gaussian noise vector with zero mean and variance $N_0 = S / SNR$. The channel matrix $\mathbf{H}$ is complex Gaussian with zero mean and unit variance, and to make the received Signal-to-Noise Ratio (SNR) $SNR$, the total transmitted power is scaled as $S$. The input-output relation in
(\ref{eq:detected_symbol}) is decomposed into two equations as
\begin{equation}
\begin{split}
\tilde{\mathbf{r}}_{k}^p = \boldsymbol{\Gamma}_p \boldsymbol{\Theta}_p \tilde{\mathbf{x}}_k^p + \tilde{\mathbf{n}}_k^p, \\
\tilde{\mathbf{r}}_{k}^n = \boldsymbol{\Gamma}_n \tilde{\mathbf{x}}_k^n + \tilde{\mathbf{n}}_k^n. \label{eq:deteced_symbol_decomposed}
\end{split}
\end{equation}

The location of the coded bit $c_{k'}$ within the complex-valued symbol sequence $\tilde{\mathbf{X}}$ is known as $k' \rightarrow (k, l, i)$, where $k$, $l$, and $i$ are the time instant in $\tilde{\mathbf{X}}$, the symbol position in $\tilde{\mathbf{x}}_k^\prime$, and the bit position on the label of the scalar symbol $\tilde{x}_{k,l}^\prime$, respectively. Let $\tilde{\chi}_{b}^{i}$ denote a subset of $\tilde{\chi}$ whose labels have $b \in \{0, 1\}$ in the $i^{th}$ bit position. By using the location information and the input-output relation in (\ref{eq:detected_symbol}), the receiver calculates the Maximum Likelihood (ML) bit metrics for $c_{k'}$ as
\begin{align}
\gamma^{l,i}(\tilde{\mathbf{r}}_{k}, c_{k'}) = \min_{\tilde{\mathbf{x}}^ \in \tilde{\xi}_{c_{k'}}^{l,i}} \| \tilde{\mathbf{r}}_{k} - \boldsymbol{\Gamma} \boldsymbol{\Theta} \tilde{\mathbf{x}} \|^2, \label{eq:ML_bit_metrics}
\end{align}
where $\tilde{\xi}_{c_{k'}}^{l,i}$ is a subset of $\tilde{\chi}^S$, defined as
\begin{align*}
\tilde{\xi}_{b}^{l,i} = \{ \tilde{\mathbf{x}} = [\tilde{x}_1 \, \cdots \, \tilde{x}_S ]^T : \tilde{x}_{s|s=l} \in \tilde{\chi}_{b}^{i}, \textrm{ and } \tilde{x}_{s|s \neq l} \in \tilde{\chi}\}.
\end{align*}
In particular, the bit metrics, equivalent to
(\ref{eq:ML_bit_metrics}) for partial precoding, are
\begin{align}
\gamma^{l,i}(\tilde{\mathbf{r}}_{k}, c_{k'}) =  \left\{
\begin{array}{ll}
\min\limits_{\tilde{\mathbf{x}} \in \tilde{\psi}_{c_{k'}}^{l,i}} \| \tilde{\mathbf{r}}_{k}^p - \boldsymbol{\Gamma}_p \boldsymbol{\Theta}_p \tilde{\mathbf{x}} \|^2, & \textrm{ if $1 \leq l \leq P$}, \\
\min\limits_{\tilde{x} \in \tilde{\chi}_{c_{k'}}^{i}} |r_{k,l} - \lambda_{l'} \tilde{x} |^2, & \textrm{ if $P+1 \leq l \leq S$},
\end{array} \right.
\label{eq:ML_bit_metrics_PPMB}
\end{align}
where $\tilde{\psi}_{b}^{l,i}$ is a subset of $\tilde{\chi}^P$, defined as 
\begin{align*}
\tilde{\psi}_{b}^{l,i} = \{ \tilde{\mathbf{x}} = [\tilde{x}_1 \, \cdots \, \tilde{x}_P ]^T : \tilde{x}_{v|v=l} \in \tilde{\chi}_{b}^{i}, \textrm{ and } \tilde{x}_{v|v \neq l} \in \tilde{\chi}\},
\end{align*}
%which is mapped from the set $\tilde{\xi}_b^{l,i}$ by a surjective function $f(\tilde{\mathbf{x}})$, for $\tilde{\mathbf{x}} = [\tilde{x}_1 \, \cdots \, \tilde{x}_S]^T$, defined as 
%\begin{equation*}
%f\left( \tilde{\mathbf{x}} \right) = [\tilde{x}_1 \, \cdots \, \tilde{x}_P]^T,
%\end{equation*}
and $l'$ is an entry in $\mathbf{b}_n$, corresponding to the subchannel mapped by $\mathbf{T}$. Finally, the ML decoder, which uses Viterbi decoding, makes decisions according to the rule
\begin{align}
\mathbf{\hat{c}} = \arg\min_{\mathbf{c}} \sum_{k'} \gamma^{l,i}(\tilde{\mathbf{r}}_{k}, c_{k'}).
\label{eq:Decision_Rule}
\end{align}

\section{Reduced Complexity Decoding for BICMB-CP} \label{sec:Decoding}

Recall that $l$ is the symbol position in $\tilde{\mathbf{x}}_k^\prime$. If $P+1 \leq l \leq S$ for (\ref{eq:ML_bit_metrics_PPMB}), the complex-valued scalar symbol carrying the coded bit $c_{k^\prime}$ is non-precoded. The non-precoded bit metric is the same as BICMB and can be decoded with low complexity using the technique presented in \cite{Akay_BICM_LCD}.

If $1 \leq l \leq P$ for (\ref{eq:ML_bit_metrics_PPMB}), the complex-valued scalar symbol carrying the coded bit $c_{k^\prime}$ is precoded. The computational complexity for the precoded bit metric is much higher than the non-precoded bit metric. Exhaustive search requires exponential complexity according to the modulation alphabet size and the dimension of the constellation precoder. The total number of lattice points needed to be searched is $|\tilde{\chi}|^{P-1}|\tilde{\chi}_{c_{k'}}^i|=\frac{|\tilde{\chi}|^P}{2}$. In this section, techniques are focus on reducing the complexity of precoded bit metrics calculation. 

\subsection{Calculating Precoded Bit Metrics By SD}
SD is used to reduce the complexity of exhaustive search by only searching lattice points inside a sphere with radius $\delta$ \cite{Jalden_SD}. Let $\tilde{\mathbf{G}}=\boldsymbol{\Gamma}_p \boldsymbol{\Theta}_p$, then SD is employed to solve
\begin{equation}
\gamma^{l,i}(\tilde{\mathbf{r}}_k, c_{k^\prime}) = \min\limits_{\tilde{\mathbf{x}} \in \tilde{\Omega}} \| \tilde{\mathbf{r}}_{k}^p - \tilde{\mathbf{G}} \tilde{\mathbf{x}} \|^2
\label{eq:SD_precoded_bit_metrics}
\end{equation}
where $\tilde{\Omega} \subset \tilde{\psi}_{c_{k'}}^{l,i}$, and $\Vert \tilde{\mathbf{r}}_{k}^p - \tilde{\mathbf{G}} \tilde{\mathbf{x}} \Vert^2<\delta^{2}$.

The $P$-dimensional complex-valued input-output relation of the precoded part in (\ref{eq:deteced_symbol_decomposed}) can be transformed into a 2P-dimensional real-valued problem \cite{Jalden_SD}:
\begin{equation}
\mathbf{r}_k^p = \mathbf{G} \mathbf{x}_k^p + \mathbf{n}_k^p, 
\label{eq:real_preocded_input_output}
\end{equation}
where $\mathbf{r}_k^p$, $\mathbf{G}$, $\mathbf{x}_k^p$, and $\mathbf{n}_k^p$ are corresponding real-valued representations of $\tilde{\mathbf{r}}_k^p$, $\tilde{\mathbf{G}}$, $\tilde{\mathbf{x}}_k^p$, and $\tilde{\mathbf{n}}_k^p$, respectively. For square QAM where $M$ is an even integer, the first and the remaining $\frac{M}{2}$ bits of labels for the $2^M$-QAM are generally Gray coded separately as two $2^{\frac{M}{2}}$-PAM constellations, and represent the real and the imaginary axes respectively. Assume that the same Gray coded mapping scheme is used for the the real and the imaginary axes. As a result, each element of $\mathbf{x}_k^p$ belongs to a real-valued signal set $\chi$, and one bit in the label of $\mathbf{x}_k^p $ corresponds to $c_{k^\prime}$. The new position of $c_{k^\prime}$ in the label of $\mathbf{x}_k^p $ needs to be acquired as $k^\prime\rightarrow(k,\hat{l},\hat{i})$, which means $c_{k^\prime}$ lies in the $\hat{i}^{th}$ bit position of the label for the $\hat{l}^{th}$ element of real-valued vector symbol $\mathbf{x}_k^p $. Let $\chi_b^{\hat{i}}$ denote a subset of $\chi$ whose labels have $b \in \{0, 1\}$ in the $\hat{i}^{th}$ bit position. Define $\psi_{c_{k'}}^{\hat{l},\hat{i}} \subset \chi^{2P}$ as 
\begin{align*}
\psi_{b}^{\hat{l},\hat{i}} = \{ \mathbf{x} = [x_1 \, \cdots \, x_{2P} ]^T : x_{v|v=\hat{l}} \in \chi_{b}^{\hat{i}}, \textrm{ and } x_{v|v \neq \hat{l}} \in \chi\}.
\end{align*}
Then (\ref{eq:SD_precoded_bit_metrics}) is rewritten as 
\begin{equation}
\gamma^{l,i}(\tilde{\mathbf{r}}_k, c_{k^\prime}) = \min\limits_{\mathbf{x} \in \Omega} \| \mathbf{r}_k^p - \mathbf{G} \mathbf{x} \|^2
\label{eq:SD_real_precoded_bit_metrics}
\end{equation}
where $\Omega \subset \psi_{c_{k'}}^{\hat{l},\hat{i}}$, and $\Vert \mathbf{r}_{k}^p - \mathbf{G} \mathbf{x} \Vert^2<\delta^{2}$.
By using the QR decomposition of $\mathbf{G}=\mathbf{QR}$, where $\mathbf{R}$ is an upper triangular matrix, and the matrix $\mathbf{Q}$ is unitary, (\ref{eq:SD_real_precoded_bit_metrics}) is rewritten as 
\begin{equation}
\gamma^{l,i}(\tilde{\mathbf{r}}_k, c_{k^\prime}) = \min\limits_{\mathbf{x} \in \Omega} \| \breve{\mathbf{r}}_k^p - \mathbf{R} \mathbf{x} \|^2 \label{eq:SD_real_QR_precoded_bit_metrics}
\end{equation}
where $\breve{\mathbf{r}}_k^p=\mathbf{Q}^H \mathbf{r}_k^p$.

SD can now be viewed as a pruning algorithm on a tree of depth $2P$, whose branches correspond to elements drawn from the set $\chi$, except for branches of the layer $u=\hat{l}$, which correspond to elements drawn from the set $\chi^{\hat{i}}_{c_{k'}}$. SD starts the search process from the root of the tree, and then searches down along branches until the total weight of a node exceeds the square of the sphere radius, $\delta^{2}$. At this point, the corresponding branch is pruned, and any path passing through that node is declared as improbable for a candidate solution. Then the algorithm backtracks, and proceeds down a different branch. Once a valid lattice point at the bottom level of the tree is found within the sphere, $\delta^{2}$ is set to the newly-found point weight, thus reducing the search space for finding other candidate solutions. In the end, the candidate solution corresponding to the path from the root to the leaf which is inside the sphere with the lowest weight is picked, and the corresponding weight is set to be the bit metric value. If no candidate solution is found, the tree will be searched again with a larger initial radius. SD can achieve the same performance as exhaustive search.

The node weight is calculated as \cite{Azzam_SD_NLR}, \cite{Azzam_SD_RLR}
\begin{equation}
w(\mathbf{x}^{(u)})=w(\mathbf{x}^{(u+1)})+w_{pw}(\mathbf{x}^{(u)}) \label{eq:node_weight}
\end{equation}
with $w(\mathbf{x}^{(2P+1)})=0$, $w_{pw}(\mathbf{x}^{(2P+1)})=0$, and $u=2P,2P-1,\cdots,1$, where $\mathbf{x}^{(u)}$ denotes the partial vector symbol at layer $u$. The partial weight $w_{pw}(\mathbf{x}^{(u)})$ is written as
\begin{equation}
w_{pw}(\mathbf{x}^{(u)})=|\breve{r}_{k,u}^p-\sum^{2P}_{v=u}{R_{u,v}x_v}|^{2} \label{eq:partial_weight}
\end{equation}
where $\breve{r}_{k,u}^p$ is the $u^{th}$ element of $\breve{\mathbf{r}}_k^p$, $R_{u,v}$ is the $(u,v)^{th}$ element of $\bf R$, and $x_v$ is the $v^{th}$ element of $\mathbf{x} \in \psi_{b}^{\hat{j},\hat{i}}$.

\subsection{Acquiring Initial Radius By ZF-DFE}
The initial radius $\delta$ should be chosen properly, so that it is not too small or too large. Too small an initial radius results in too many unsuccessful searches and thus increases complexity, while too large an initial radius results in too many lattice points to be searched.

In this work, for $c_{k'}=b$ where $b \in \{0,1\}$, ZF-DFE is used to acquire a estimated real-valued vector symbol $\breve{\mathbf{x}}_k^b$, which is also the Baiba point \cite{Agrell_CPS}. Then the square of initial radius $\delta_b^2$, which guarantees no unsuccessful searches is calculated by
\begin{equation}
\delta_b^2=\Vert \breve{\mathbf{r}}_k^p-\mathbf{R}\breve{\mathbf{x}}_k^b \Vert^2.
\label{eq:IR_ZFDFE}
\end{equation}

The estimated real-valued vector symbol $\breve{\mathbf{x}}_k^b$ is detected successively starting from $\breve{x}_{k, 2P}^b$ until $\breve{x}_{k, 1}^b$, where $\breve{x}_{k, u}^b$ denotes the $u^{th}$ element of $\breve{\mathbf{x}}_k^b$. The decision rule on $\breve{x}_{k, u}^b$ is
\begin{equation}
\breve{x}_{k, u}^b=\left\{
\begin{array}{ll}
\arg\min\limits_{x\in\chi} {| \breve{r}_{k,u}^p-\sum^{2P}_{v=u+1}{R_{u,v}\breve{x}_{k, v}^b}-R_{u,u}x |},&u\neq \hat{l},\\ \arg\min\limits_{x\in\chi^{\hat{i}}_b} {| \breve{r}_{k,u}^p-\sum^{2P}_{v=u+1}{R_{u,v}\breve{x}_{k, v}^b}-R_{u,u}x |},&u=\hat{l}.\\
\end{array}
\right.
\label{eq:ZFDFE} 
\end{equation}

The estimation of the symbols (\ref{eq:ZFDFE}) can be carried out recursively by rounding (or quantizing) to the nearest constellation element in $\chi$ or $\chi^{\hat{i}}_b$.

\subsection{Reducing Number of Executions in SD}

For the $k^{th}$ time instant, the precoded real-valued vector symbol $\mathbf{x}_k^p$ carries $MP$ bits. Since each bit generates two bit metrics for $c_{k'}=0$ and $c_{k'}=1$, then $2MP$ precoded bit metrics in total need to be acquired. However, some precoded bit metrics have the same value, hence SD can be modified to be executed less than $2MP$ times, as mentioned in \cite{Studer_SOSD}.

Define $\hat{\mathbf{x}}_k$, $\hat{\mathbf{x}}_k^{c_{k'}}$, and $\gamma_{k}$ as
\begin{equation}
\hat{\mathbf{x}}_k = \arg\min\limits_{\mathbf{x} \in \chi^{2P}} \| \breve{\mathbf{r}}_k^p - \mathbf{R} \mathbf{x} \|^2,
\label{eq:ML_symbol} 
\end{equation}
\begin{equation}
\hat{\mathbf{x}}_k^{c_{k'}} = \arg\min\limits_{\mathbf{x} \in \psi_{c_{k'}}^{\hat{l},\hat{i}}} \| \breve{\mathbf{r}}_k^p - \mathbf{R} \mathbf{x} \|^2,
\label{eq:bit_metric_symbol_estimated} 
\end{equation}
and
\begin{equation}
\gamma_{k} = \| \breve{\mathbf{r}}_k^p - \mathbf{R} \hat{\mathbf{x}}_k \|^2,
\label{eq:ML_symbol_weight} 
\end{equation}
respectively. 
%Then 
%\begin{equation}
%\gamma^{l,i}(\tilde{\mathbf{r}}_k, c_{k^\prime}) =  \| \breve{\mathbf{r}}_k^p - \mathbf{R} \hat{\mathbf{x}}_k^{c_{k'}} \|^2,
%\label{eq:bit_metric_symbol} 
%\end{equation}
Note that $\psi_0^{\hat{l},\hat{i}} \cup \psi_1^{\hat{l},\hat{i}} = \chi^{2P}$ and $\psi_0^{\hat{l},\hat{i}} \cap \psi_1^{\hat{l},\hat{i}} = \emptyset$. Then
\begin{equation}
\gamma_{k} = \min{\{\gamma^{l,i}(\tilde{\mathbf{r}}_k, c_{k^\prime}=0), \gamma^{l,i}(\tilde{\mathbf{r}}_k, c_{k^\prime}=1) \} },
\label{eq:symbol_weight_relation} 
\end{equation}
which means that, for the $MP$ bits corresponding to $\mathbf{x}_k^p$, the smaller precoded bit metric for each bit of $c_{k'}=0$ and $c_{k'}=1$ have the same value $\gamma_{k}$. 

Let $\hat{b}_{\hat{i}}^{\hat{l}}\in \{0,1\}$ denotes the value of the $\hat{i}^{th}$ bit in the label of $\hat{x}_{k,\hat{l}}$, which is the $\hat{l}^{th}$ element of $\hat{\mathbf{x}}_k$. Then  
\begin{equation}
\gamma^{l,i}(\tilde{\mathbf{r}}_k, c_{k^\prime}=\hat{b}_{\hat{i}}^{\hat{l}})=  \gamma^k.
\label{eq:bit_metric_smaller} 
\end{equation}

First, two bit metrics $\gamma^{l,i}(\tilde{\mathbf{r}}_k, c_{k^\prime}=0)$ and $\gamma^{l,i}(\tilde{\mathbf{r}}_k, c_{k^\prime}=1)$ for one of the $MP$ bits corresponding to $\mathbf{x}_k^p$ and their related  $\hat{\mathbf{x}}_k^{c_{k'}}$ are derived by SD. Then the $\hat{\mathbf{x}}_k^{c_{k'}}$ corresponding to the smaller bit metric is chosen to be $\hat{\mathbf{x}}_k$, and $\gamma_{k}$ is acquired by (\ref{eq:symbol_weight_relation}). For each of the other $MP-1$ bits, $\gamma^{l,i}(\tilde{\mathbf{r}}_k, c_{k^\prime}=\hat{b}_{\hat{i}}^{\hat{l}})$ is acquired by (\ref{eq:bit_metric_smaller}), and $\gamma^{l,i}(\tilde{\mathbf{r}}_k, c_{k^\prime}=\bar{\hat{b}}_{\hat{i}}^{\hat{l}})$ is calculated by SD. Consequently, the execution number of SD for one time instant is reduced from $2MP$ to $MP+1$.

\subsection{Reducing Number of Operations in SD}

In our previous work \cite{Li_RCSD_arXiv}, a technique was introduced to implement SD with low computational complexity, which achieves the same performance as exhaustive search. The technique in this paper can be employed to achieve substantial further complexity reduction for BICMB-CP. In this subsection, a brief description of the technique is presented for reducing the number of real multiplications.

Note that for one channel realization, both $\mathbf{R}$ and $\chi$ are independent of time. In other words, to decode different received symbols for one channel realization, the only term in (\ref{eq:partial_weight}) which depends on time is $r_{k,u}^p$. Consequently, a check-table $\mathbb{T}$ is constructed to store all terms of $R_{u,v}x$, where $R_{u,v}\neq0$ and $x \in \chi$, before starting the tree search procedure. Equations (\ref{eq:node_weight}) and (\ref{eq:partial_weight}) imply that only one real multiplication is needed by using $\mathbb{T}$ instead of $2P-u+2$ for each node to calculate the node weight. As a result, the number of real multiplications can be significantly reduced. 

Note that $\chi$ can be divided into two smaller sets $\chi_{1}$ with negative elements and $\chi_{2}$ with positive elements. Any negative element in $\chi_{1}$ has a positive element with the same absolute value in $\chi_{2}$. Consequently, in order to build $\mathbb{T}$, only terms of $R_{u,v}x$, where $R_{u,v}\neq0$ and $x\in\chi_{1}$, need to be calculated and stored. Since the channel is assumed to be flat fading, only one $\mathbb{T}$ needs to be built in one burst. If the burst length is very long, its complexity can be neglected.

In our previous work \cite{Azzam_SD_NLR}, \cite{Azzam_SD_RLR}, a new lattice representation was introduced. In this work, the same lattice representation is employed to (\ref{eq:real_preocded_input_output}) but with a new application. The structure of the lattice representation becomes advantageous after applying the QR decomposition to $\mathbf{G}$. By doing so, and due to the special form of orthogonality between each pair of columns, all elements $R_{u,u+1}$ for $u=1,3,\ldots,2P-1$ in the upper triangular matrix $\mathbf{R}$ become zero. The locations of these zeros introduce orthogonality between the real and the imaginary parts of every detected symbol, which can be taken advantage of to reduce the computational complexity of SD. 

Based on this feature, SD is modified in the following way: once the tree is searched in layer $u$, where $u$ is an odd number, partial weights of this node and all of its brother nodes are computed, temporally stored, and recycled when calculating partial node weights with the same grandparent node of layer $u+2$ but with different parent nodes of layer $u+1$. By implementing the modification, further complexity reduction is achieved. 

\section{Simulation Results} \label{sec:Results}
Since the $P$-dimensional complex-valued input-output relation of the precoded part in (\ref{eq:deteced_symbol_decomposed}) can be viewed as a $P$-dimensional BICMB-FP, BICMB-FP is considered to verify the proposed technique. Exhaustive Search (EXH), Conventional SD (CSD), and Proposed Smart Implementation (PSI) which combines Section \ref{sec:Decoding}.C and Section \ref{sec:Decoding}.D, are applied. The average number of real multiplications, the most expensive operations in terms of machine cycles, for acquiring one bit metric is calculated at different SNR. %Since the reductions in complexity are substantial, they are expressed as orders of magnitude (in approximate terms) in the sequel.

Fig. \ref{fig:2x2} shows comparisons for $2\times2$ $S=2$ $R_c=\frac{2}{3}$ BICMB-FP. For $4$-QAM, the complexity of EXH is reduced by $0.4$ and $0.5$ orders of magnitude at low and high SNR respectively, by CSD. PSI yields larger reductions by $1.1$ and $1.2$ orders of magnitude at low and high SNR respectively. In the case of $64$-QAM, reductions between CSD and EXH are $1.5$ and $2.1$ orders of magnitude at low and high SNR respectively, while larger reductions of $2.6$ and $3.0$ are achieved by PSI.

Similarly, Fig. \ref{fig:4x4} shows complexity comparisons for $4\times4$ $S=4$ $R_c=\frac{4}{5}$ BICMB-FP. For $4$-QAM, the complexity of EXH decreases by $1.3$ and $1.5$ orders of magnitude at low and high SNR respectively. PSI gives larger reductions by $2.3$ orders of magnitude at low SNR, and $2.4$ orders of magnitude at high SNR. For the $64$-QAM case, reductions between EXH and CSD by $3.2$ and $4.4$ orders of magnitude are observed at low and high SNR respectively, while larger reductions by $4.4$ and $5.4$ are achieved by PSI.

Simulation results show that CSD reduces the complexity substantially compared to EXH, and the complexity can be further reduced significantly by PSI. The reductions become larger as the constellation precoder dimension and the modulation alphabet size increase. One important property of our decoding technique needs to be emphasized is that the substantial complexity reduction achieved causes no performance degradation. 

%Moreover, the proposed smart implementation can be applied to any convolutional coded MIMO system.

\ifCLASSOPTIONonecolumn
\begin{figure}[!m]
\centering
\scalebox{.7}{\includegraphics{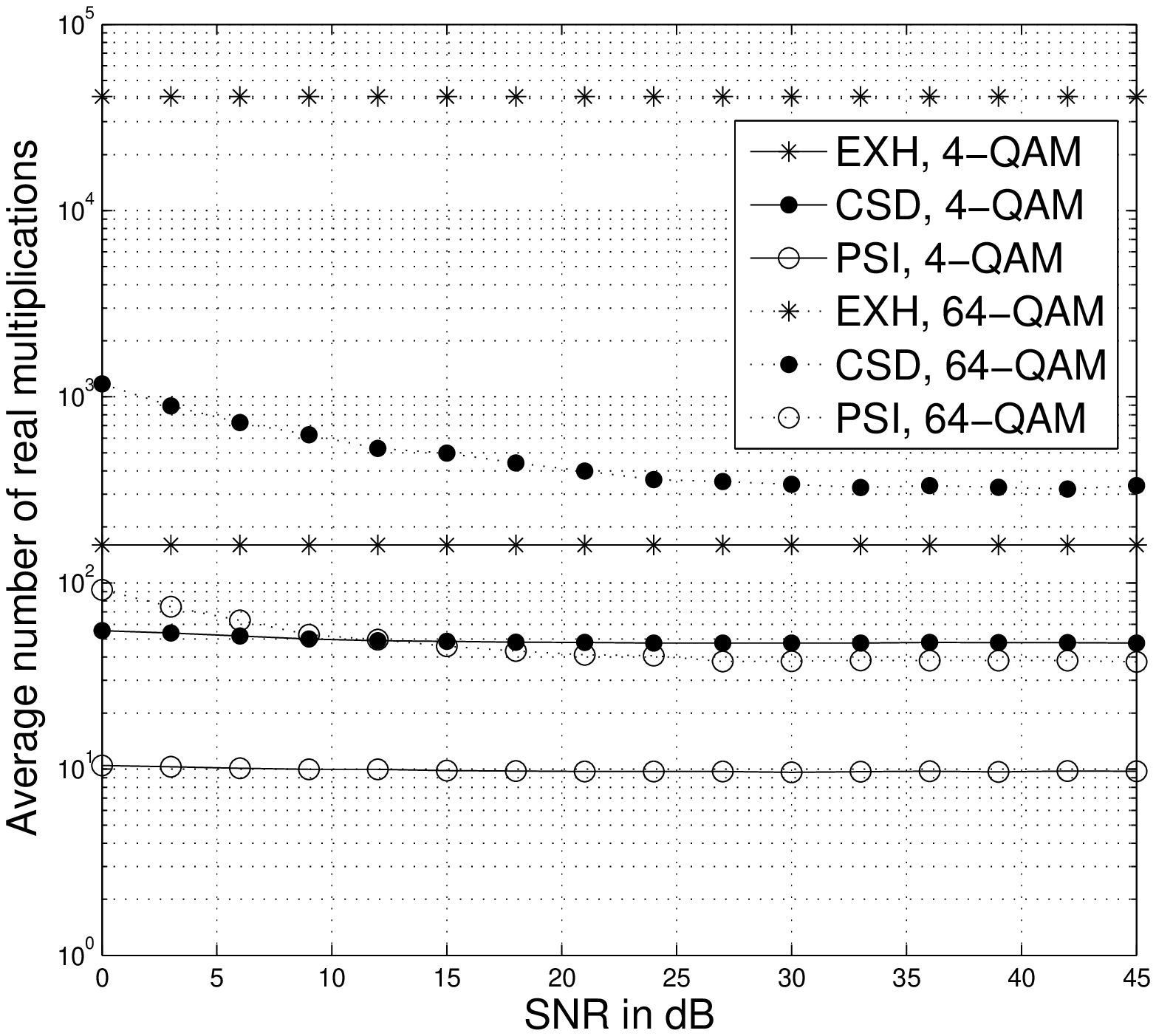}}
\caption{Average number of real multiplications vs. SNR for $2\times2$ $S=2$ BICMB-FP.}
\label{fig:2x2}
\end{figure}

\begin{figure}[!m]
\centering
\scalebox{.7}{\includegraphics{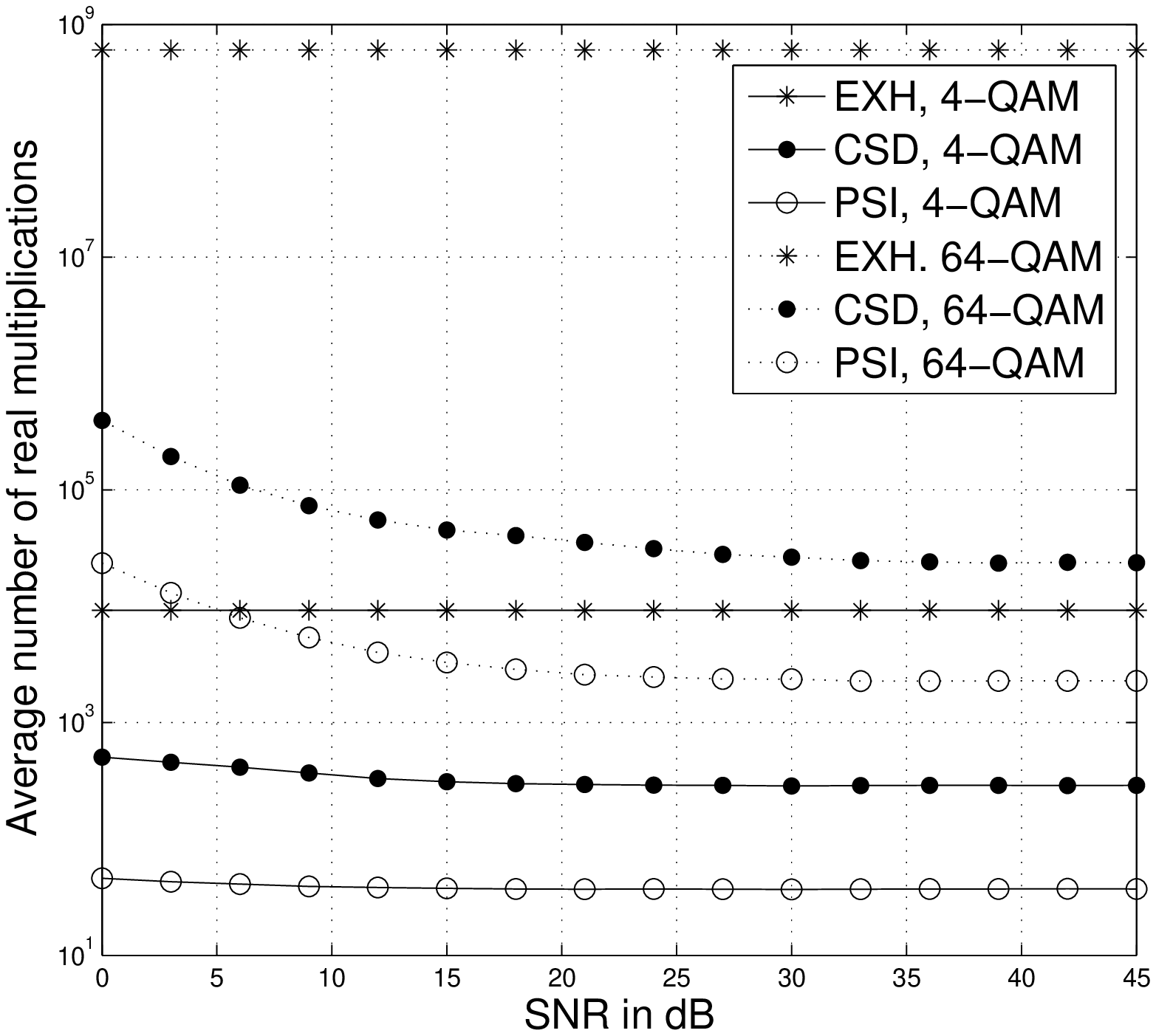}}
\caption{Average number of real multiplications vs. SNR for $4\times4$ $S=2$ BICMB-FP.}
\label{fig:4x4}
\end{figure}

\else
\begin{figure}[!t]
\centering
\scalebox{.5}{\includegraphics{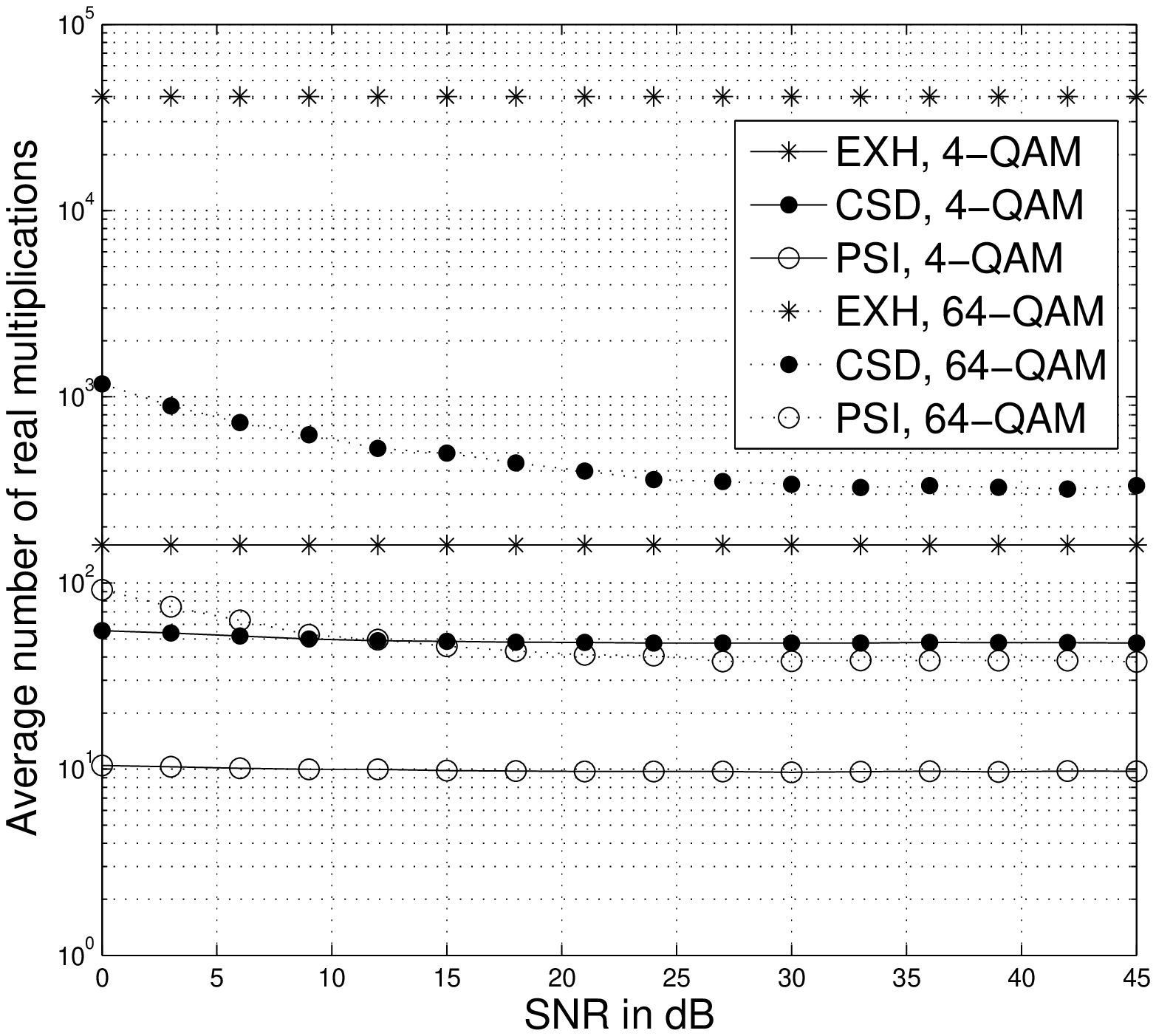}}
\caption{Average number of real multiplications vs. SNR for $2\times2$ $S=2$ BICMB-FP.}
\label{fig:2x2}
\end{figure}

\begin{figure}[!t]
\centering
\scalebox{.5}{\includegraphics{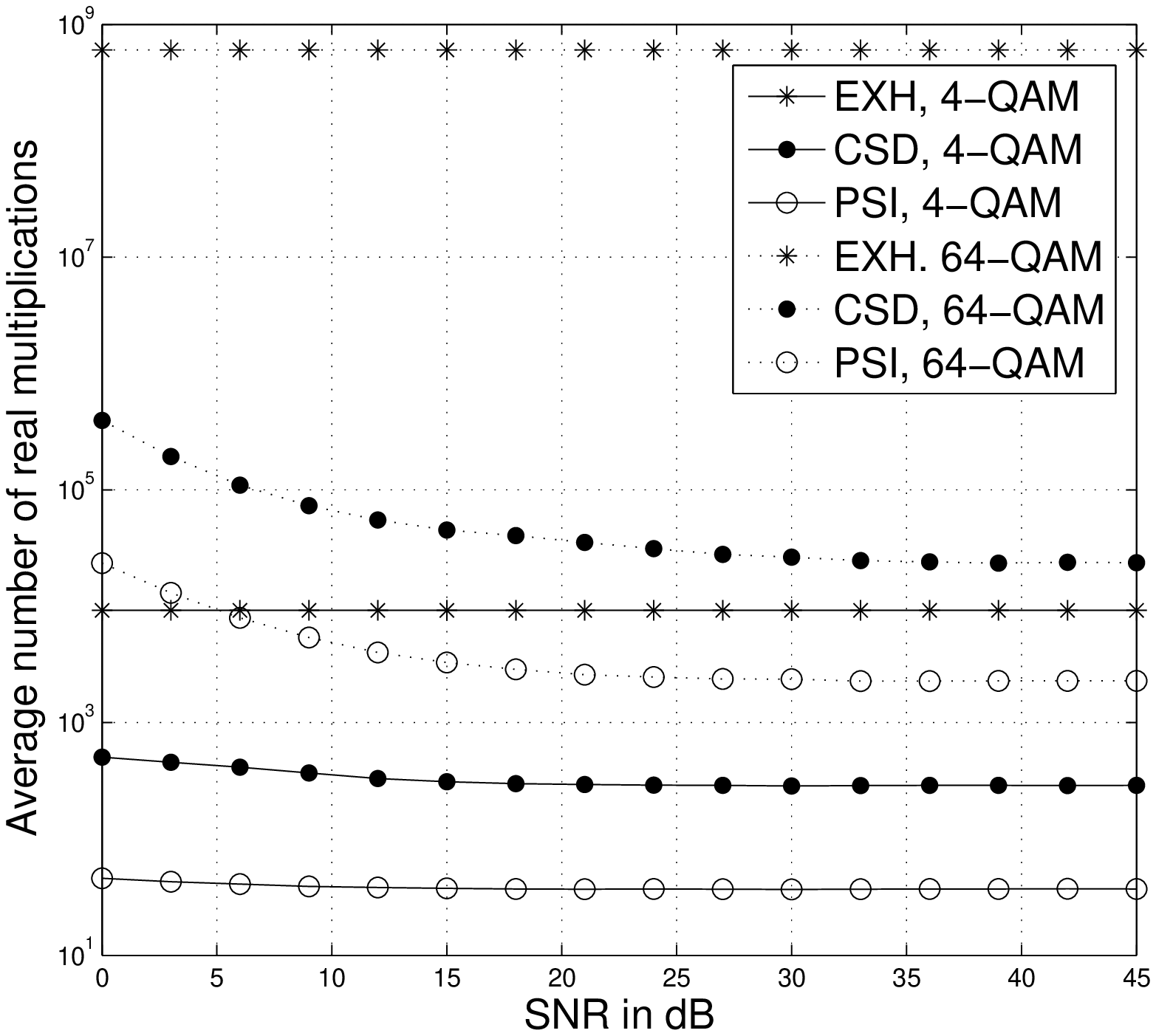}}
\caption{Average number of real multiplications vs. SNR for $4\times4$ $S=2$ BICMB-FP.}
\label{fig:4x4}
\end{figure}

\fi

\section{Conclusion} \label{sec:Conclusion}

In this paper, a reduced complexity decoding scheme for BICMB-CP is presented. SD with initial radius calculated by ZF-DFE is used to acquire precoded bit metrics needed for the Viterbi decoder. SD can achieve the same performance as exhaustive search, and more importantly, achieves a substantial complexity reduction. Two techniques are applied to reduce both the number of executions and operations for SD substantially. Therefore, BICMB-CP can be considered as a practical application for MIMO systems requiring high throughput with the full diversity order. The reduced complexity decoding in this paper can be applied to any convolutional coded MIMO system.

\ifCLASSOPTIONonecolumn \newpage \fi

\bibliographystyle{IEEEtran}
\bibliography{IEEEabrv,Mybib}

\begin{thebibliography}{10}
\providecommand{\url}[1]{#1}
\csname url@samestyle\endcsname
\providecommand{\newblock}{\relax}
\providecommand{\bibinfo}[2]{#2}
\providecommand{\BIBentrySTDinterwordspacing}{\spaceskip=0pt\relax}
\providecommand{\BIBentryALTinterwordstretchfactor}{4}
\providecommand{\BIBentryALTinterwordspacing}{\spaceskip=\fontdimen2\font plus
\BIBentryALTinterwordstretchfactor\fontdimen3\font minus
  \fontdimen4\font\relax}
\providecommand{\BIBforeignlanguage}[2]{{%
\expandafter\ifx\csname l@#1\endcsname\relax
\typeout{** WARNING: IEEEtran.bst: No hyphenation pattern has been}%
\typeout{** loaded for the language `#1'. Using the pattern for}%
\typeout{** the default language instead.}%
\else
\language=\csname l@#1\endcsname
\fi
#2}}
\providecommand{\BIBdecl}{\relax}
\BIBdecl

\bibitem{Paulraj_ST}
A.~Paulraj, R.~Nabar, and D.~Gore, \emph{Introduction to Space-Time Wireless
  Communication}.\hskip 1em plus 0.5em minus 0.4em\relax Cambridge University
  Press, 2003.

\bibitem{Zheng_DM}
L.~Zheng. and D.~Tse, ``{Diversity and Multiplexing: a Fundamental Tradeoff In
  Multiple-antenna Channels},'' \emph{{IEEE} Trans. Inf. Theory}, vol.~49,
  no.~5, pp. 1073--1096, May 2003.

\bibitem{Jafarkhani_STC}
H.~Jafarkhani, \emph{Space-Time Coding: Theory and Practice}.\hskip 1em plus
  0.5em minus 0.4em\relax Cambridge University Press, 2005.

\bibitem{Palomar_JTRBD}
D.~P. Palomar, J.~M. Cioffi, and M.~A. Lagunas, ``{Joint Tx-Rx Beamforming
  Design for Multicarrier MIMO Channels: A Unified Framework for Convex
  Optimization},'' \emph{{IEEE} Trans. Signal Process.}, vol.~51, no.~9, pp.
  2381--2401, Sep. 2003.

\bibitem{Sengul_DA_SMB}
E.~Sengul, E.~Akay, and E.~Ayanoglu, ``{Diversity Analysis of Single and
  Multiple Beamforming},'' \emph{{IEEE} Trans. Commun.}, vol.~54, no.~6, pp.
  990--993, Jun. 2006.

\bibitem{Akay_BICMB}
E.~Akay, E.~Sengul, and E.~Ayanoglu, ``{Bit-Interleaved Coded Multiple
  Beamforming},'' \emph{{IEEE} Trans. Commun.}, vol.~55, no.~9, pp. 1802--1811,
  Sep. 2007.

\bibitem{Akay_On_BICMB}
\BIBentryALTinterwordspacing
E.~Akay, H.~J. Park, and E.~Ayanoglu. (2008) {On "Bit-Interleaved Coded
  Multiple Beamforming"}. arXiv: 0807.2464. [Online]. Available:
  \url{http://arxiv.org}
\BIBentrySTDinterwordspacing

\bibitem{Park_DA_BICMB}
H.~J. Park and E.~Ayanoglu, ``{Diversity Analysis of Bit-Interleaved Coded
  Multiple Beamforming},'' in \emph{Proc. IEEE ICC 2009}, Dresden, Germany,
  Jun. 2009.

\bibitem{Park_BICMB_CP}
------, ``{Bit-Interleaved Coded Multiple Beamforming with Constellation
  Precoding},'' in \emph{Proc. IEEE ICC 2010}, Cape Town, South Africa, May
  2010.

\bibitem{Han_SD_IR}
H.~G. Han, S.~K. Oh, S.~J. Lee, and D.~S. Kwon, ``{Computational Complexities
  of Sphere Decoding According to Initial Radius Selection Schemes and an
  Efficient Initial Radius Reduction Scheme},'' in \emph{Proc. IEEE GLOBECOM
  2005}, vol.~4, St. Louis, MO, USA, Nov. 2005, pp. 2354--2358.

\bibitem{Agrell_CPS}
E.~Agrell, T.~Eriksson, A.~Vardy, and K.~Zeger, ``{Closest Point Search in
  Lattices},'' \emph{{IEEE} Trans. Inf. Theory}, vol.~48, no.~8, pp.
  2201--2214, Aug. 2002.

\bibitem{Haccoun_PCC}
D.~Haccoun and G.~Begin, ``{High-Rate Punctured Convolutional Codes for Viterbi
  and Sequential Decoding},'' \emph{{IEEE} Trans. Commun.}, vol.~37, no.~11,
  pp. 1113--1125, Nov. 1989.

\bibitem{Akay_BICM_LCD}
E.~Akay and E.~Ayanoglu, ``{Bit-Interleaved Coded Modulation: Low Complexity
  Decoding},'' in \emph{Proc. VTC 2004}, vol.~1, Milan, Italy, May 2004, pp.
  328--332.

\bibitem{Jalden_SD}
J.~Jald{\'{e}}n and B.~Ottersten, ``{On the Complexity of Sphere Decoding in
  Digital Communications},'' \emph{{IEEE} Trans. Signal Process.}, vol.~53,
  no.~4, pp. 1474--1484, Apr. 2005.

\bibitem{Azzam_SD_NLR}
L.~Azzam and E.~Ayanoglu, ``{Reduced Complexity Sphere Decoding for Square QAM
  via a New Lattice Representation},'' in \emph{Proc. IEEE GLOBECOM 2007},
  Washington, D.C., USA, Nov. 2007, pp. 4242--4246.

\bibitem{Azzam_SD_RLR}
------, ``{Reduced Complexity Sphere Decoding via a Reordered Lattice
  Representation},'' \emph{{IEEE} Trans. Commun.}, vol.~57, no.~9, pp.
  2564--2569, Sep. 2009.

\bibitem{Studer_SOSD}
C.~Studer, A.~Burg, and H.~B{\"{o}}lcskei, ``{Soft-Output Sphere Decoding:
  Algorithms and VLSI Implementation},'' \emph{{IEEE} J. Sel. Areas Commun.},
  vol.~26, no.~2, Feb. 2008.

\bibitem{Li_RCSD_arXiv}
\BIBentryALTinterwordspacing
B.~Li and E.~Ayanoglu. (2009) {Reduced Complexity Sphere Decoding}.
  arXiv:0909.0555v2. [Online]. Available: \url{http://arxiv.org}
\BIBentrySTDinterwordspacing

\end{thebibliography}
\end{document}